\documentclass[aps,pre,showpacs,preprint]{revtex4-1}
\usepackage[english,UKenglish]{babel}
\usepackage{amsmath, amsthm, amsfonts}
\usepackage[total={17cm,21cm},centering]{geometry}
\usepackage{graphicx}
\usepackage{makeidx}
\usepackage{color}
\usepackage[table,xcdraw]{xcolor}
\usepackage{float}
\usepackage{subfigure}
\usepackage{listings}
\usepackage{bm}
\usepackage[final]{pdfpages}
\usepackage[labelfont=bf]{caption}

\begin{document}

\title{Entropy geometric construction of a pure substance with normal, superfluid and supersolid phases}
\author{Manuel Mendoza-L\'opez and V\'{\i}ctor Romero-Roch\'{\i}n} 
\affiliation{Instituto de F\'{\i}sica. Universidad Nacional Aut\'onoma de M\'exico.\\
Apartado Postal 20-364, 01000 M\'exico, Cd.Mx., Mexico}
\date{\today}

\begin{abstract}
\indent Using the laws of thermodynamics together with empirical data, we present a qualitative geometric construction of the fundamental relation of a pure substance $S = S(E,N,V)$, with $S$ entropy, $E$ energy, $N$ number of particles and $V$ volume. We analyze two very general type of substances, a ``normal" and a ``quantum" one,  the main difference between them being that the latter presents superfluid phases. It is found that the constant entropy level curves are completely different in both cases, in the normal substances being obtuse while acute in quantum ones. A concomitant signature of the previous result is that the chemical potential can be both positive and negative in quantum substances, but only negative in normal ones. Our results suggest the existence of a region in the quantum substances that may be identified as a supersolid phase. We also make emphasis on the relevance of the present study within the context of superfluidity in ultracold gases.
\end{abstract}

\maketitle

\section{Introduction}

While the elucidation of equations of state and susceptibilities of pure substances, both theoretically and experimentally, is vigorously pursued due to their immediate relevance in basic and applied research, rarely one finds analysis of fundamental thermodynamic relations which, if known, would allow for the knowledge in turn of all thermodynamic properties \cite{LL,Callen}. Indeed, if getting hold of equations of state and susceptibilities, such as isothermal expansions and specific heats, is already difficult, the construction of a fundamental relation appears as an even harder endeavor. Nevertheless, this can be done qualitatively if, on the one hand one knows empirical data in the form of measured phase diagrams and, on the other, one makes use of the power of the laws of thermodynamics. This is not just an academic exercise. As we shall see, the fundamental relation indicates features that are absent in usual phase diagrams, such as $p - T$ diagrams with $p$ pressure and $T$ temperature. These features allows us to pose restrictions both in theoretical models as well as in measurable quantities. Furthermore, the fundamental relation may also indicate the existence of phases, with their own associated phenomena, that have not been measured or not looked for. \\

The fundamental relation that we address in this article is the entropy $S$ as a function of the extensive variables energy $E$, number of particles $N$ and volume $V$. However, since entropy is also extensive, it can be written as $S = V s(e,n)$ where $s = S/V$, $e= E/V$ and $n = N/V$, namely, volume densities of entropy, energy and number of particles. Therefore, the task we perform is the geometric construction of $s = s(e,n)$, a single valued surface, function of two variables only, $e$ and $n$. As we review below, as dictated by thermodynamics, the function $s$ is not only single valued, it is also concave in both variables $e$ and $n$. The difficulty, as we shall see, resides in the correct localization of the different phases. The reward of this construction, as we have already mentioned, is that the knowledge of $s = s(e,n)$ completely suffices to determine {\it all} the thermodynamic properties of the substance. \\

In Section II we present generic empirical $p-T$ phase diagrams, based on the measured phase diagrams of Ar and $^4$He \cite{MedicionesArgon, MedicionesHelio1, MedicionesHelio2, MedicionesHelio3}. The generic substance based on Ar will be termed ``normal", while the one related to $^4$He will be called ``quantum" due to the presence of the superfluid phase. An ultimate understanding of both diagrams certainly requires the use of quantum mechanics but it is the nature of superfluidity that suggested to us the use of those adjectives. The main purpose of  Section II is to indicate the main features of the phases as well as the order of the different phase transitions. We take this as ``empirical" data on which our further analysis will be based.\\

Section III is devoted to a brief review of the laws of thermodynamics and their implications on the geometric and topological characteristics of the surface $s = s(e,n)$. In particular, we shall consider systems whose energy spectra is unbounded from above, since the molecular kinetic energy is always present in the considered systems and, therefore, the temperature will be considered positive always, $T > 0$ \cite{LL}. Strictly speaking, $T = 0$ is banned by the Third Law. Further, we shall make an strong emphasis that although, in principle, the chemical potential $\mu$ can be expressed with respect to an arbitrary reference, we can fix this reference by imposing that in the limit of arbitrarily low particle densities, $n \to 0$, the substance must obey the thermodynamics of a classical ideal gas. This will give a clear meaning to the sign of the chemical potential.\\

Sections IV and V discuss in detail the construction of the surfaces $s = s(e,n)$ for the normal and the quantum substances respectively. We will emphasize the crucial fact that, within the convention used, the chemical potential is always negative for normal substances while it can change sign in the quantum one. This will also be reflected in the geometric characteristics of the entropy level surfaces, being obtuse in the normal case and acute in the quantum version. As an important consequence, it will be shown that the superfluid phase occurs mostly in the region of positive chemical potential and, as we will further argue, there appears a region in the solid phase with positive chemical potential which may be identified as a supersolid ``phase" \cite{Ho,Supersolidez1,Supersolidez2,Supersolidez3,Supersolidez4}. We conclude in Section VI with a set of Final Remarks.\\

\section{Thermodynamic phenomenology of a pure substance}

A pure substance, composed of a sole chemical compound,  can be  found in the form of solid, liquid, gas, superfluid or in phase coexistence of the former. We consider as typical examples of pure substances as H$_2$O, Ar, CO$_2$, $^4$He, $^3$He, among others, of which an abundant experimental data exist. Gravitational, electric and magnetic effects will not be considered in the thermodynamic description of a pure substance here. \\

For the sake of simplicity, we classify the pure substances according to their quantum characteristics into normal and quantum substances. Microscopic descriptions of normal and quantum substances assume intermolecular interaction potentials, typically pairwise. Normal substances can be observed in solid, liquid and gas phases. Additionally, due to particular qualities of the intermolecular potential \cite{MedicionesHelio1, MedicionesHelio2, MedicionesHelio3}, quantum substances can be observed in one more phase, namely, the superfluid phase. 
That is, we denote as ``quantum" substances to the aforementioned by the fact that superfluidity is a macroscopic quantum phenomenon, described primarily in the context of many-body quantum theory, see e.g. Ref. \cite{Fetter}. There are other systems that can be called quantum substances as well, for example, a Fermi gas, a Mott insulator, or even a magnet, but here we shall mainly be concerned with the shared thermodynamic characteristics of $^4$He, $^3$He, and the recent ultracold alkali vapors, both bosonic, such as $^{87}$Rb, $^{23}$Na, $^7$Li  \cite{Cornell,Ketterle,Hulet,Pethick}, and fermionic, such as $^6$Li and $^{40}$K \cite{OHara,Jin,Levin}.  \\

\begin{figure}[t]
\centering
\hfill
\subfigure[Argon]{\includegraphics[width=8cm,height=8cm]{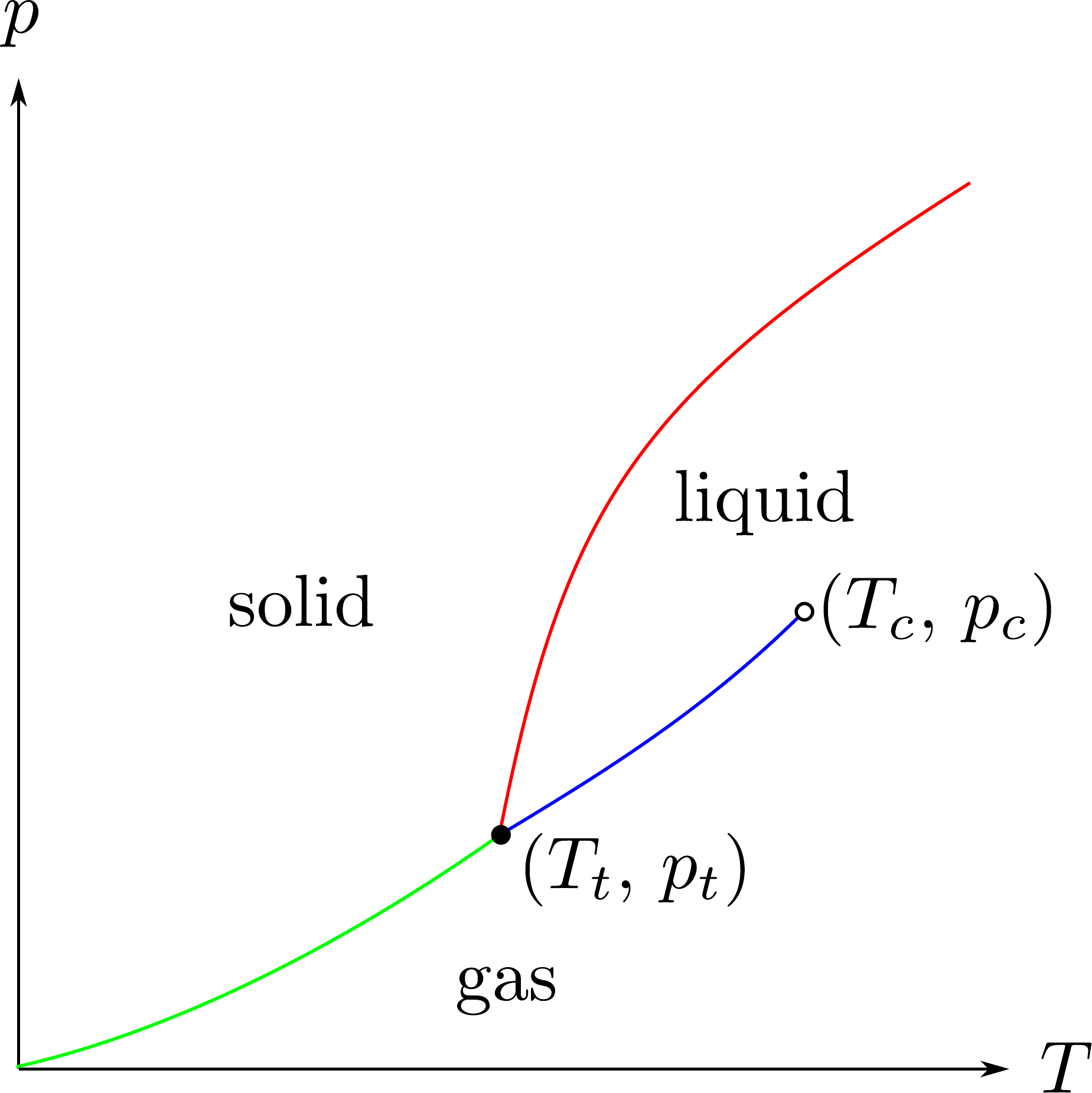}}
\hfill
\subfigure[Helium-4]{\includegraphics[width=8cm,height=8cm]{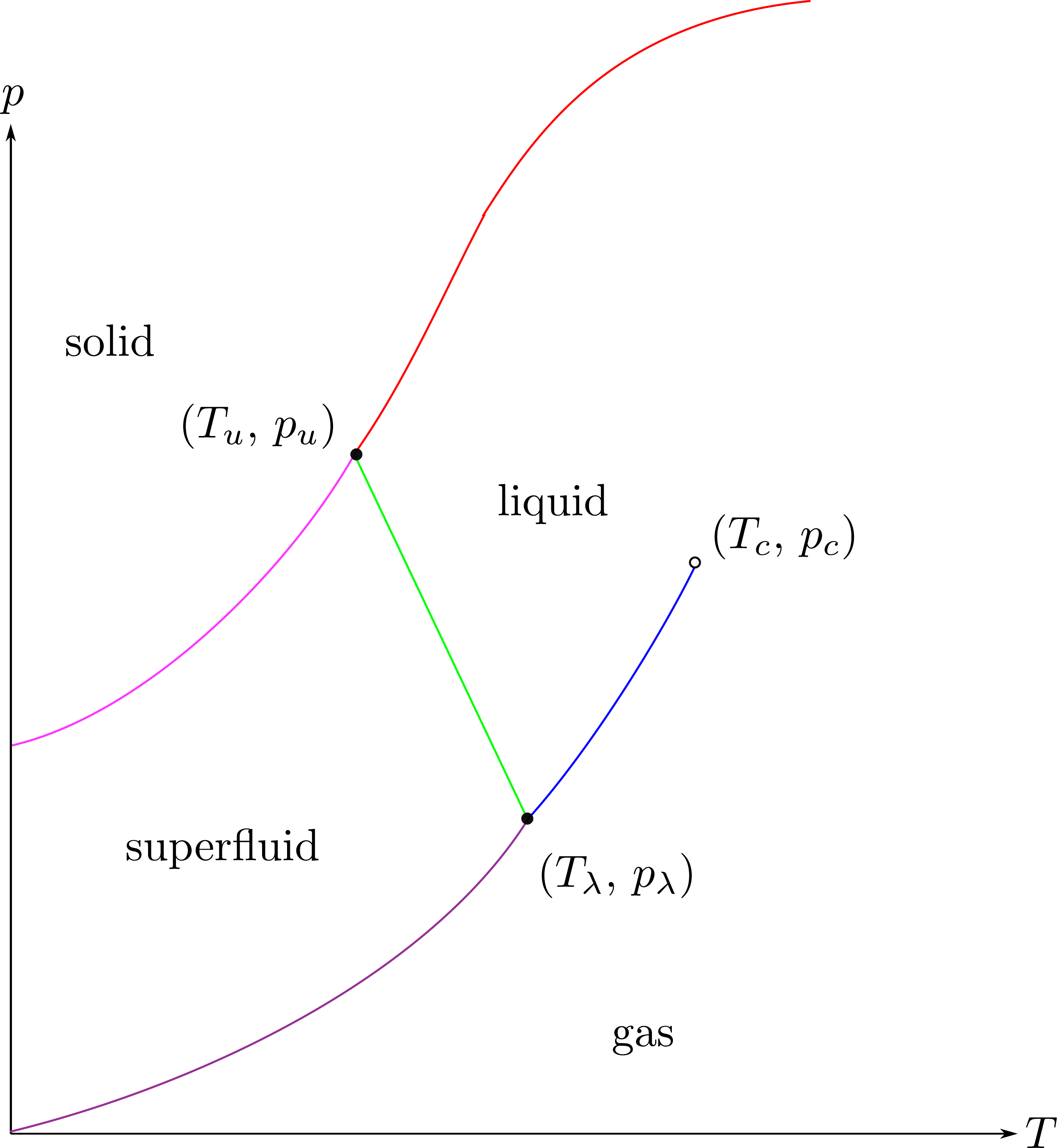}}
\hfill
\caption{(Color online) Generic $p-T$ phase diagrams for normal (a) and quantum (b) substances. Diagrams are not drawn to scale. According to Ref. \cite{MedicionesArgon}, the coordinates of the triple (solid black) and the critical (empty black) points of Argon are ($T_t=83.804$ K, $p_t=0.06895$ MPa) and ($T_c=150.6633$ K, $p_c=4.860$ MPa), respectively. In the case of $^4$He, there are three isolated points: two of them (solid black) are the ends of the $\lambda$-line, which is the line separating the liquid and the superfluid phases; the other is the critical point (empty black). From Refs. \cite{MedicionesHelio1, MedicionesHelio2, MedicionesHelio3}, the coordinates of these points are ($T_{u}=1.7633$ K, $p_{u}=3.01340$ MPa), ($T_{\lambda}=2.1720$ K, $p_{\lambda}=0.00504$ MPa) and ($T_c=5.2014$ K, $p_c=0.22746$ MPa), respectively. The point of the highest temperature in the $\lambda$-line is called the $\lambda$-point, and the point of the lowest temperature could be called the upper-triple point \cite{MedicionesHelio2, MedicionesHelio3} because it could represent thermodynamic equilibrium between the superfluid, solid and liquid phases.}
\label{DiagramaPVST}
\end{figure}

Figure \ref{DiagramaPVST} shows generic $p-T$ diagrams of (a) a normal substance and (b) a quantum one. Although stylized for didactical purposes, those diagrams are based on the experimental phase diagrams of Ar and $^4$He \cite{MedicionesArgon, MedicionesHelio1, MedicionesHelio2, MedicionesHelio3}.  According to the Gibbs phase rule, areas in a phase diagram correspond to a unique phase in equilibrium and lines to the coexistence between two phases. Isolated points in the diagram corresponds either to a triple point or to a critical point. In the figure caption we specify the actual values of all those relevant pressures and temperatures. As we will explicitly do so in Sections IV and V, these phase diagrams will allow us to construct the fundamental relationship $s = s(e,n)$. \\

It is well known that for a normal substance, with the exception of the critical point that is a continuous second order phase transition, all the coexistence curves represent discontinuous first order phase transitions. For the quantum substance, it is also known \cite{MedicionesHelio1, MedicionesHelio2, MedicionesHelio3, LineaFusion} that the liquid-solid, liquid-gas and superfluid-solid are first order phase transitions. The critical point and the $\lambda-$line liquid-superfluid are second order phase transitions. Although for $^4$He there are references, such as Refs. \cite{Zemansky,Huang,Fisher}, that suggest that the coexistence curve gas-superfluid is first order, there are no experimental data confirming this is the case. On the other hand, all the recent experimental and theoretical work on the fermionic and bosonic degenerate ultracold gases, that suffer a gas-superfluid transition \cite{Pethick,Dalfovo}, indicate that such a transition is similar to the $\lambda$ transition and, therefore, that it is second order. Thus, without affecting our overall conclusions, we shall also consider that such a transition is indeed second order.\\

\section{Thermodynamic laws and their restrictions on $s = s(e,n)$}

Besides the assumed fact that thermodynamics describes equilibrium states only, it is very important to emphasize at the outset that a given value of the set of variables $(E,N,V)$ of a pure substance uniquely specifies an equilibrium state. Therefore, the entropy function $S(E,N,V) = Vs(e,n)$ is a single valued function of its variables. In the same fashion, any  Legendre transform of $S(E,N,V)$, say $V$ by $p$ or $N$ by $\mu$, is also a single valued fundamental relation bearing the same information just in the transformed variables \cite{LL,Callen}. Here, we concentrate on the relation $s = s(e,n)$.\\

The First Law of Thermodynamics is,
\begin{equation}
ds=\frac{1}{T}de-\frac{\mu}{T}dn,
\label{PrimeraLey}
\end{equation}
which indicates that
\begin{equation}
\frac{1}{T} =\left(\frac{\partial s}{\partial e} \right)_{n} \>\>\>\>\>{\rm and} \>\>\>\>\>\frac{\mu}{T}=-\left(\frac{\partial s}{\partial n} \right)_{e} .\label{Tmu}
\end{equation}
The extensitivity of $S = V s(e,n)$ implies that the pressure is given by,
\begin{equation}
p = Ts - e + \mu n .
\end{equation}

The Second Law of Thermodynamics asserts that the entropy $s = s(e,n)$ is a concave function of its variables, namely, 
\begin{equation}
\left( \frac{\partial ^2 s}{\partial e^2}\right)_n < 0
\label{Condicion1}
\end{equation}
and 
\begin{equation}
\left( \frac{\partial ^2 s}{\partial n^2}\right)_e - \frac{\left(\frac{\partial ^2 s}{\partial e \partial n}\right)^2}{\left(\frac{\partial ^2 s}{\partial e^2}\right)_n} < 0.
\label{Condicion2}
\end{equation}

These inequalities establish the stability of the thermodynamic states \cite{LL,Callen}. The first condition is equivalent to stating that the specific heat at constant volume $c_v$ is positive, whereas the second one that the isothermal compressibility $\kappa_T$ is also positive, 
\begin{equation}
c_v = T \left(\frac{\partial s}{\partial T}\right)_v > 0 \>\>\>\>{\rm and} \>\>\>\> \kappa_T =  \frac{1}{n} \left(\frac{\partial n}{\partial p}\right)_T > 0 .\label{suscep}
\end{equation}
When any of these conditions is not met, the system suffers a phase transition to enforce stability.\\

Phase transitions are classified into first and second order. In a first order phase transition the extensive quantities $s$, $e$, and $v$ are discontinuous while the intensive ones $T$, $p$ and $\mu$ are always continuous.  In a second order phase transition all the variables are continuous but the 
remarkable characteristic of these transitions is that, typically, the specific heat and the isothermal compressibility become infinite at the transition. The main lesson from this behavior is that the function $s = s(e,n)$ is not analytic at critical points \cite{Fisher,Ma,Amit}. For the purposes of our geometric construction we shall use, with no loss of generality, that since $\kappa_T \to \infty$ at critical points, this implies that $(\partial p / \partial n)_T \to 0$ at those points. That is, the isothermal  curves, $T =$ constant, $p$ vs $n$  become flat at second order phase transitions but remain monotonically increasing at any other thermodynamic state.\\

Since we are dealing with systems whose energy spectra are unbounded from above, negative temperatures are ruled out \cite{LL,Ramsey,VRR-NT} and, therefore, the Third Law can simply be expressed as the requirement that the temperature is always positive, $T > 0$. Zero temperature can be considered as a limit only and, as we shall see, this will also be reflected in the form of the surface $s = s(e,n)$. Due to the identification of temperature by  the first of Eqs. (\ref{Tmu}), this implies that the isochores curves, $n = $ constant, $s$ vs $e$ are monotonously increasing with their slope becoming infinite at $T \to 0$. \\

A most important issue is the sign of the chemical potential $\mu$. On the one hand, there is no thermodynamic law that bears an implication on such an issue. At most, the Second Law requires the surface $s = s(e, n)$ to be concave but this does not imply that the isoenergetic curves, $e = $ constant, $s$ vs $n$ have to be monotonously increasing, see Eqs. (\ref{Tmu}) and (\ref{Condicion1}). On the other hand, it is commonly believed that the chemical potential can only be defined up to an arbitrary constant since the origin of the energy can also be arbitrarily fixed. This constant, however, can indeed be fixed for atomic substances by imposing that in the limit of a very dilute system, $n \to 0$, the system approaches an ideal gas whose energy can be taken as definite positive since it arises from the atomic kinetic energy only. From a thermodynamic point of view, we can impose that in the normal gas region of the phase diagram, as $n \to 0$, the entropy of the gas must approach \cite{LL}
\begin{equation}
s(e,n) \to  nk \left\{ \ln \frac{1}{n} \left(\frac{4 \pi m}{3 h^2} \frac{e}{n} \right)^{3/2} + \frac{5}{2} \right\} \label{Sideal}
\end{equation}
which is the classical ideal gas expression for a system of $n$ non interacting atoms of mass $m$. In the above expression $k$ and $h$ are Boltzmann and Planck constants. This restriction also fixes the pressure to become $p \to nkT$ and the energy $e \to  (3/2) nkT$ in the  limit $n \to 0$ for fixed temperature. Clearly, with this condition, the energy of the gas phase is limited to have positive values only. Because of the identification of the chemical potential by the second of Eqs. (\ref{Tmu}), and because $T > 0$, the reference of the chemical potential is also fixed, becoming negative in the mentioned limit $n \to 0$. \\

To look further into the sign of the chemical potential recall that, according to statistical physics, the entropy is given by \cite{LL},
\begin{equation}
S(E,N,V) = k \ln \Omega(E,V,N)
\end{equation}
where $\Omega(E,V,N)$ is the number of states of the system for given values of $(E,V,N)$. The thermodynamic limit is assumed to  be taken.  Thus, for a fixed value of the energy per volume $e = E/V =$ constant, a negative sign of the chemical potential implies that the number of states increases as the particle density $n = N/V$ also increases, see Eq. (\ref{Tmu}). Conversely, a positive $\mu$ signals that the number of states {\it decreases} when particle density $n$ increases. The first behavior is the common one in classical systems since, even if the energy per particle $e/n$ diminishes, the fact that there are more particles implies a larger number of available states. Of course, as the number of particles increases even further, the increase of entropy will still do so but at a slower rate. The extreme case of positive chemical potential implies that as the energy per particle decreases the number of available states also decreases. This implies a kind of  ``condensation" that can only be explained within the use of quantum mechanics. This is the common behavior of ideal Bose and Fermi gases \cite{LL,Pathria}, as well of those weakly interacting ones, as given by BCS \cite{Fetter,BCS} and Bogoliubov \cite{Pethick,Bogoliubov} theories: all of these do show a positive chemical potential. The reason behind such a behavior is that, as the entropy decreases with even larger densities, there exists a limiting value of $n$ for which the entropy becomes zero, $s = 0$. That is, for the given fixed value of $e$, such an extreme value of $n$ corresponds to the quantum ground state of the system. \\

Because classical ideal gases, models such as van der Waals, and ideal quantum gases present negative chemical potentials before quantum effects set in (such as Bose-Einstein condensation and Fermi degeneracy \cite{LL,Pathria}), we shall make a working hypothesis that, as it will be shown, it is a correct one, namely, that in the gas phase the chemical potential is always negative. As we will see, this fact will distinctly shape the surfaces $s = s(e,n)$ for normal and quantum substances.\\

\section{Entropy construction of a generic normal substance}

\indent The logical stages in the construction of the phase diagrams that represent the function $s = s(e,n)$ of the normal and the quantum substance are the same. First, we assume the phase diagram $p-T$ as given, as shown in Fig. \ref{DiagramaPVST}. Then, we map the corresponding isotherms onto the $p-n$ diagram. These can also be checked with available experimental data \cite{MedicionesArgon} and with the predictions of van der Waals model. Third, we use the behavior of the density of the system in the $p-n$ diagram as a guide to infer the structure of the $s-n$ diagram. The very last stage consists in mapping the isoenergetic curves  in the $s-n$ diagram onto corresponding ones in the $s-e$ diagram. The last two diagrams, $s-n$ and $s-e$, can be summarized in a $n-e$ diagram with isentropic curves $s = $ constant. \\

\begin{figure}[]
\centering
\includegraphics[scale=0.35]{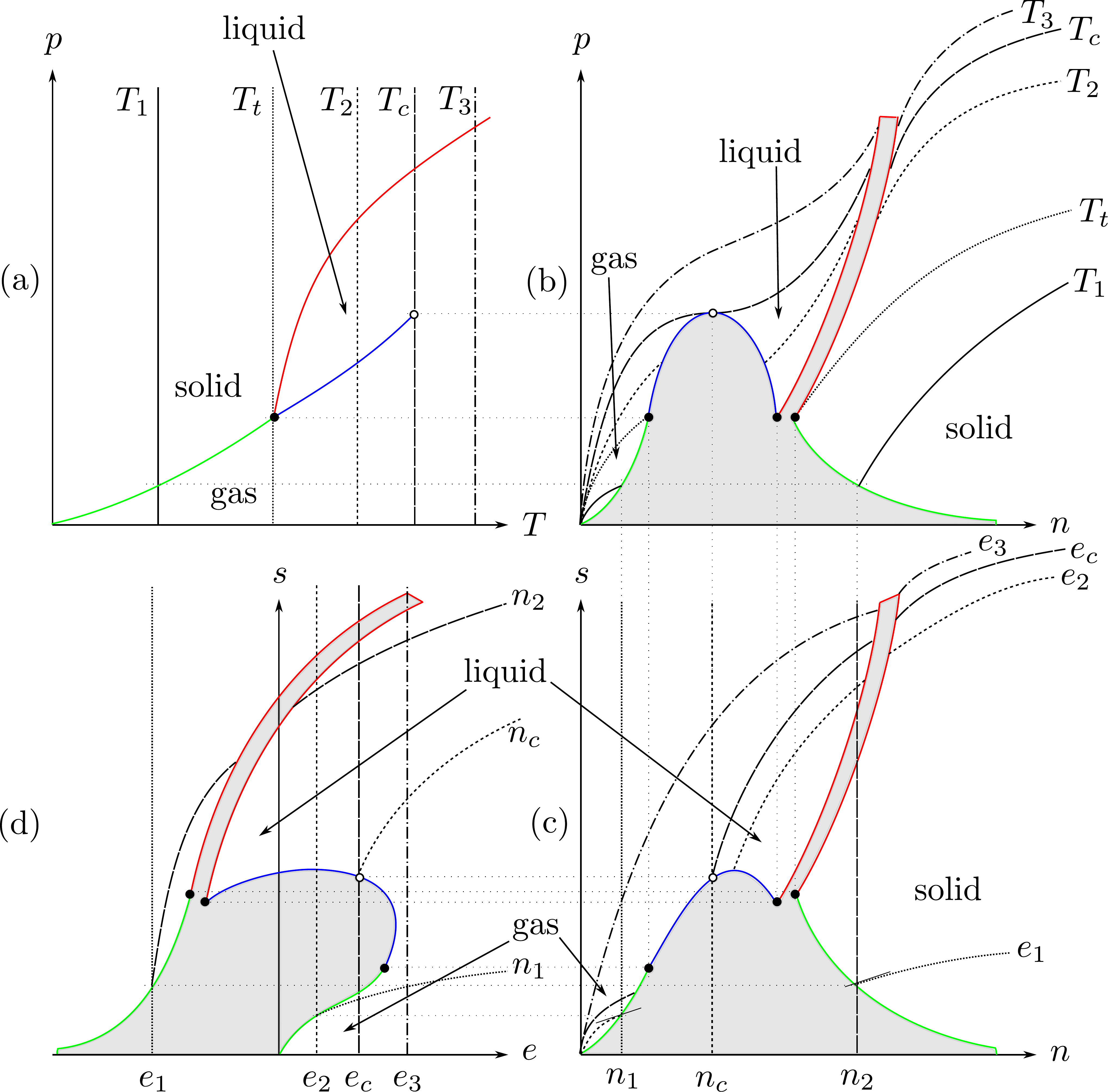}
\caption{(Color online) Generic phase diagrams for a normal substance. (a) The $p-T$ diagram is based on empirical data \cite{MedicionesArgon}. (b) We infer the structure of the $p-n$ diagram from the structure of $p-T$ and also from empirical data. The isotherms are qualitative and must obey the linearity of the ideal gas equation in the dilute regime. Note that the critical isotherm is flat at the critical point. (c) The $s-n$ diagram is built accordingly to the behaviour of the density and of the coexistence regions in the $p-n$ phase diagram. (d) Then, we obtain the $s-e$ diagram from the previous step. In both (c) and (d), we plot the same isochores $n =$ constant and isoenergetics $e = $ constant.}
\label{diagramaNormal}
\end{figure}

As described above, we start with the $p-T$ diagram of a normal substance in Fig. \ref{DiagramaPVST}(a). Consider an isothermal curve $T = $ constant. This is represented by a vertical line that can cross one or two coexistence lines, see Fig. \ref{diagramaNormal}(a) where 5 typical isotherms are plotted. Fig. \ref{diagramaNormal}(b) shows the corresponding diagram $p-n$. This diagram is also well known, as can be found in usual thermodynamics textbooks \cite{Zemansky}, but it can also be constructed from the $p-T$  diagram and the restrictions from the laws of thermodynamics. We highlight several aspects. First, the isotherms have the same pressure at coexistence. Second, we empirically set the slope of the isotherms at coexistence being higher in the denser state. And third,  the solid phase is assumed to asymptotically reach zero pressure at  infinite density $n$, consistently with the presumption that such a solid can occur at zero temperature only and in coexistence with a gas at zero density, namely, an ``unatainable" state. The most important aspect of this figure, but also a well known result, is the existence of {\it unstable} regions where no equilibrium states occur. These are the shaded areas in the figures. \\

The unstable states actually do not ``exist" and, therefore, we cannot assign a value for any of the variables $s$, $e$, $n$, etc, in those regions. This is very important for the construction of the phase diagram $s-n$ and, ultimately, for the $s-e$ one. That is, from diagram $p-n$ we can construct $s-n$, Fig. \ref{diagramaNormal}(c), using as guides the coexistence regions. As can be checked from models, such as van der Waals, and from experimental data \cite{MedicionesArgon}, at coexistence the entropy {\it per} volume $s = S/V$ is higher in the denser coexistence region; if one considers entropies per mole, the inequalities are inverted. Thus, in the triple point the entropy per volume $s$ is lowest in the gas phase and highest in the solid one. The task here is to trace isoenergetic curves $e =$ constant in the diagram $s-n$. We have two initial restrictions, first, those curves must be concave and, second, they cannot cross since $e = e(s,n)$ is also a fundamental relation \cite{Callen}. And here it comes the fundamental issue that the slope of the isochores equals $-\mu/T$, with $T > 0$. The immediate consequence is that the isoenergetics at coexistence, while they correspond to different values of energy, must have the same slope, since $\mu$ and $T$ must be continuous. Now, since we are assuming that at very low densities the ideal gas is always reached, the chemical potential starts as negative and, therefore, the isoenergetics start concave but with positive slope at $n = 0$. The question is whether the isoenergetics reach the coexistence lines with positive or negative slope. Again, from van der Waals model, one can check that chemical potential is negative in the gas-liquid coexistence states. However, for the gas-solid coexistence states one does not have a reliable model. Nevertheless, from the illustration of Fig. \ref{diagramaNormal}(c), we find that because of the existence of the unstable region and because the saturated gas curve must be much steeper than the saturated solid one, the slope of the isoenergetics at the solid-gas coexistence states, in order to have the same value, must have positive slope, namely negative chemical potential, otherwise they would either point into the unstable region or they would eventually cross. The extraordinary conclusion is that, for a normal substance the isoenergetic curves are all concave and monotonously increasing functions $s$ of $n$. That is, the chemical potential is always negative for normal substances. As we will see below, the ``intrusion" of the superfluid phase into the unstable gas-solid region makes it necessary the appearance of positive chemical potentials and, thus, the whole structure of the surface $s = s(e,n)$ for a quantum substance will be completely different than in the normal ones.\\

From the diagram $s-n$, Fig. \ref{diagramaNormal}(c), we can now construct the $s-e$ diagram, as shown in Fig. \ref{diagramaNormal}(d). Here, we have at the outset two strong restrictions, the isochores, $n = $ constant, $s$ vs $e$ must not only be concave but because their slope is the inverse of the temperature, see Eq.(\ref{Tmu}), they must be monotonic. The only delicate point is that, since we assume that for arbitrary temperature one can always lower the density to extreme small values, the energy in this limit must reach $e \sim n kT$. This implies that the gas phase must end at zero energy, setting thus its origin. This right away implies that the low entropy solid phase must have negative energy. However, nothing prevents the solid to have positive energy at higher entropies, or the liquid phase to have negative energies either. It is therefore not very complicated to conclude that the $s-e$ diagram, with the help of diagram $s-n$, appears as such. In this diagram, we have plotted several isochores to match the corresponding points values in the $s-n$ diagram. \\

\begin{figure}[]
\centering
\includegraphics[scale=0.5]{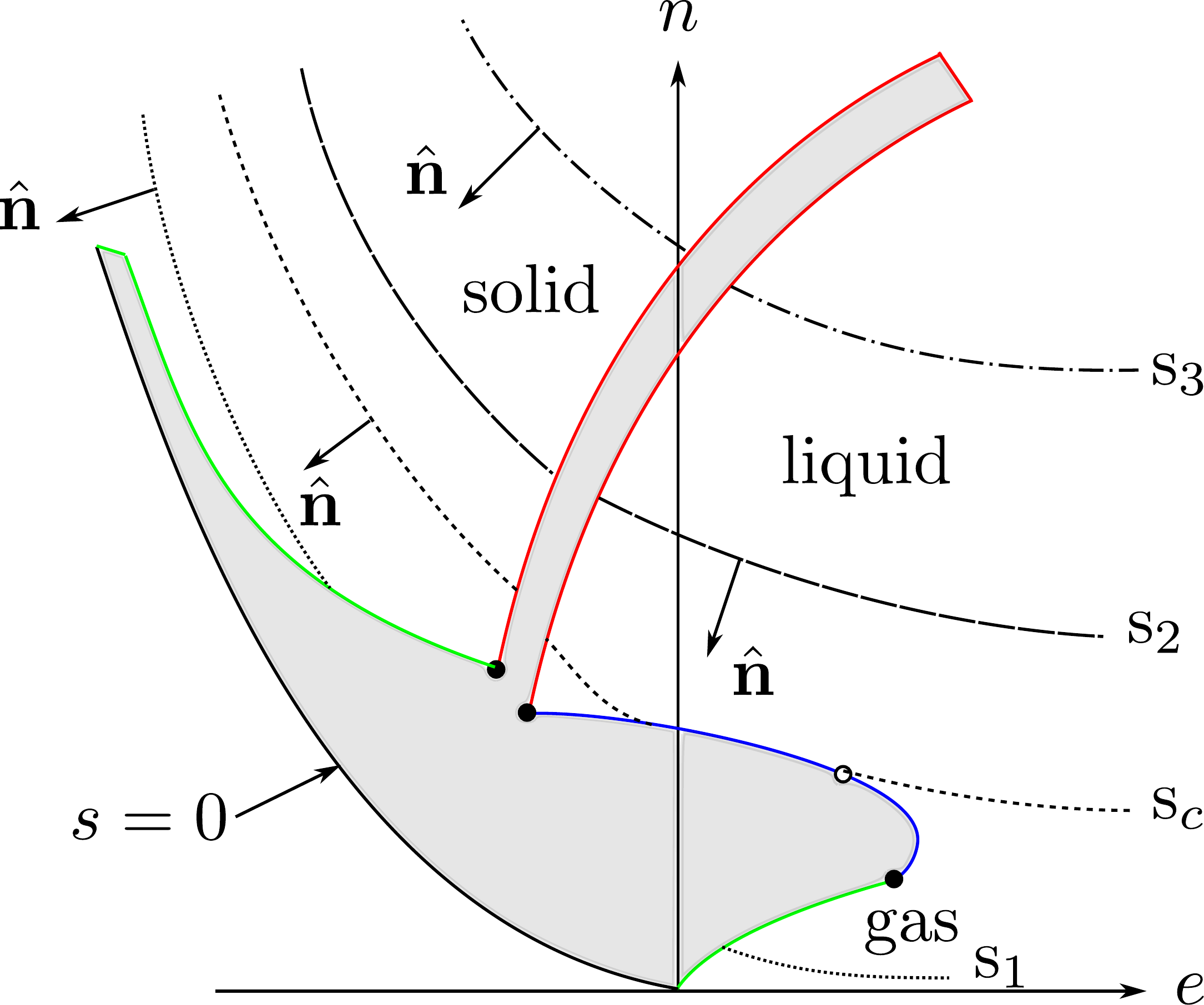}
\caption{(Color online) Isentropic level curves $s = $ constant in a $n-e$ plane. Here $s_1 < s_c < s_2 < s_3$ are drawn qualitatively. Note that in an adiabatic compression or expansion the energy is monotonous. Moreover, the tangent asymptotes to the isentropic curves,  $(n \to \infty,e \to -\infty)$ and $(n \to 0,e \to \infty)$, form an angle greater than $\frac{\pi}{2}$, namely, the adiabatic curves are all obtuse. Note also that the normal vectors have all negative components.}
\label{DiagramaNVSENormal}
\end{figure}

The diagrams $s-n$ and $s-e$ define the surface $s = s(e,n)$. In order to gain a bit more insight on such a surface, we can draw isentropic ($s = $ constant) level curves in a $n-e$ diagram. For this, we first calculate the normal vectors to the surface, placing energy $e$ along $x$, $n$ along $y$ and $s$ along $z$, in conventional cartesian coordinates. Using dimensionless units, say using the gas-liquid critical point as reference, the normal unit vectors of the surface $s = s(e,n)$ are given by,
\begin{equation}
\hat{\mathbf{n}} = \frac{1}{\sqrt{\mu^2+T^2+1}}(-1,\mu,T).
\label{normal}
\end{equation}
Thus, along the adiabatic curves $s(e,n) =$ constant, the normal components in the plane $n-e$, namely, $-1/(\mu^2+T^2+1)^{1/2}$ along $e$ and $\mu/(\mu^2+T^2+1)^{1/2}$ along $n$, always point in the negative directions because $\mu < 0$. Thus, the isentropic, or adiabatic, curves are all {\it obtuse}. This is illustrated in Fig. \ref{DiagramaNVSENormal}, where we plot several level curves of $s = $ constant. The extreme curve $s = 0$, which is degenerated with $T = 0$, is reached with infinite slope in the gas phase, with zero density, and it is never reached in the solid phase. This property of the isentropic curve indicates that in any adiabatic process the energy monotonically either increases or decreases, and the temperature always increases or decreases as well, properties that can experimentally be checked.\\

Although surfaces plotted in 2D-planes never do justice to the richness of a 3D object, one can gain further information using the normal vector $\hat {\bf n}$ given by Eq.(\ref{normal}). The main aspect we want to highlight is that, while the surface $s = s(e,n)$ is certainly smooth, it cannot be covered continuously with the same surface, that is, it is actually formed by patches of surfaces, with ``holes" on it, representing the inexistent unstable states. At any point on the surface one can set a local tangent plane whose normal is defined by the temperature $T$ and the chemical potential $\mu$, as given by Eq. (\ref{normal}). Two different coexistent states at a first order phase transition, say, $s_1 = s(e_1,n_1)$ and $s_2 = s(e_2,n_2)$ have parallel tangent planes, which yield equal temperature $T$ and chemical potential $\mu$, and their precise location is further determined by the fact that they also have the same pressure, namely, $p = e_1 - T s_1 + \mu n_1$ and $p = e_2 - T s_2 + \mu n_2$. This makes the ``sewing" of the surfaces of the different phases. The critical point is peculiar since the coexisting planes become closer and closer and finally merge in the same plane at the critical point. Thus, it seems, the surface $s = s(e,n)$ must be flat at the critical point. This represents a ``suture" of two different surfaces and the matching cannot be made analytically in general. This flatness of the surfaces at the critical point is an indication of the divergence of the specific heat and of the isothermal compressibility.

\section{Entropy construction of a generic quantum substance}

\begin{figure}[]
\centering
\includegraphics[scale=0.25]{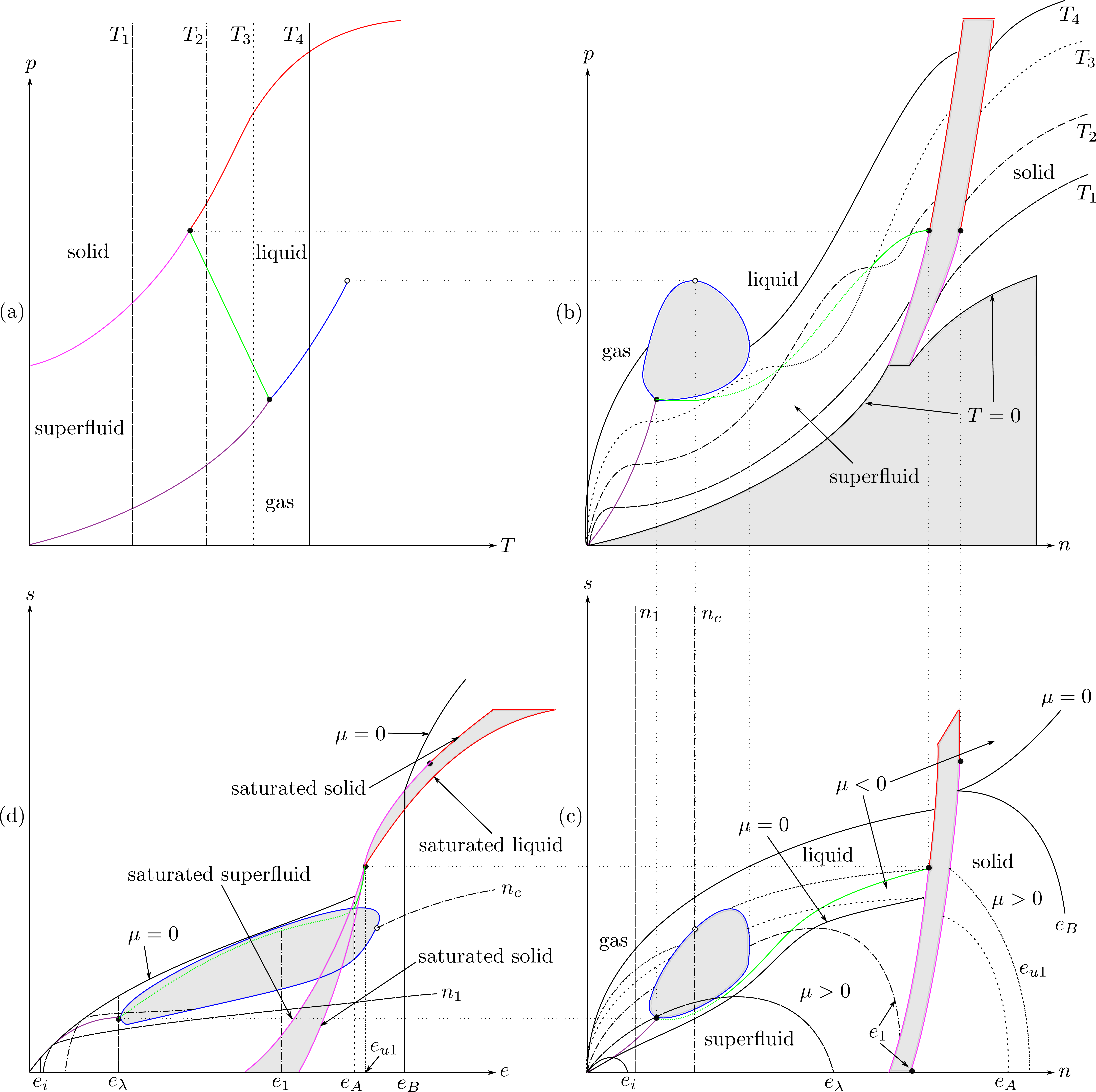}
\caption{(Color online) Generic phase diagrams for a quantum substance. (a) $p-T$ diagram based on empirical data \cite{MedicionesHelio1, MedicionesHelio2, MedicionesHelio3}. (b) $p-n$ diagram based on the $p-T$ one. The states below the line $T = 0$ are non-existent. Note that the isotherms cross each other in the vicinity of the $\lambda$-line, see text for explanation. (c) $s-n$ diagram built based on the behavior of the density and the coexistence regions in the $p-n$ diagram. The isoenergetic curves are all non-monotonic. (d) $s-e$ diagram based on the previous diagram. In both (c) and (d), we plot the same isochores $n =$ constant and isoenergetics $e = $ constant.}
\label{quantumall}
\end{figure}

\begin{figure}[h]
\centering
\includegraphics[scale=0.5]{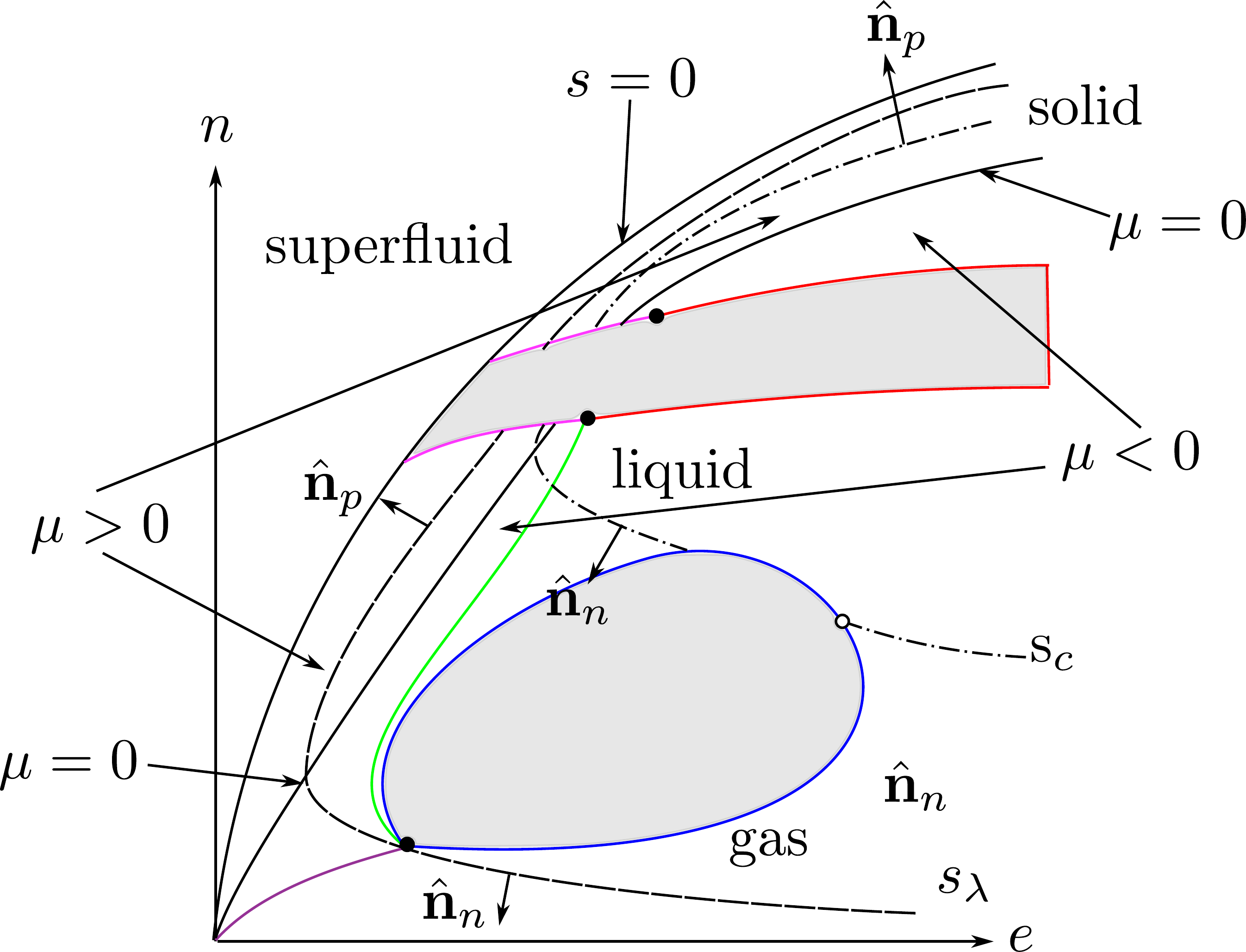}
\caption{(Color online) Adiabatics in a quantum $n$ vs $e$ phase diagram. We draw two types of normal vectors, those with negative components, and those whose components have different sign. Thus, the adiabatic processes showed, expansion or compression, have thermodynamic states with $\mu < 0$ ($\hat{\mathbf{n}}_n$), and another with $\mu > 0$ ($\hat{\mathbf{n}}_p$). The qualitative asymptotes corresponding to $n \to \infty$ and $e \to \infty$, and $n \to 0$ and $e \to \infty$ make an angle less than $\frac{\pi}{2}$.} 
\label{s-e-n-cuantico}
\end{figure}

Analogous to the previous section, the goal here is the geometric construction of the surface $s=s(e,n)$ for a quantum substance whose generic $p-T$ phase diagram is given in Fig. \ref{DiagramaPVST}(b). As mentioned before, the main characteristics of this diagram follow the properties of $^4$He, see Refs. \cite{Kapitza,AllenMisener,London,Tisza,Landau}. The procedure, in principle, is the same as for a normal substance. From the $p-T$ diagram we construct the $p-n$ diagram, then the $s-n$ one, and finally $s-e$, as illustrated in Fig. \ref{quantumall}. The last two are the corresponding projections of the surface $s = s(e,n)$. From the latter, we can plot isentropic curves in the $n-e$ plane to illustrate that for a quantum substance those curves are ``acute", see Fig. \ref{s-e-n-cuantico}. As we shall discuss, a section of the solid phase shows ``quantum" properties that makes it a candidate for the supersolid phase \cite{Ho,Supersolidez1,Supersolidez2,Supersolidez3,Supersolidez4}.\\

Regarding the $p-T$ diagram in Fig. \ref{quantumall}(a), we recall once more that the coexistence curves solid-superfluid and solid-liquid are first order. The same for the liquid-vapor curve terminating in a critical point. The $\lambda$-line, separating liquid-superfluid, is a line of second order phase transitions. This transition has been extensively studied, both experimentally and theoretically, and it has been concluded that it belongs to the 3D XY-model universality class with critical exponents $\alpha = - 0.0127$ \cite{Lipa,Ahlers1,Ahlers2} and $\nu = 0.6717$ \cite{Machta}. The usual difficulty with this transition is that while the corresponding order parameter is the complex condensate wavefunction, its conjugate field is unknown \cite{Fisher}. This prevents from obtaining the critical behavior of all thermodynamic properties across the transition. Nevertheless, there is no doubt that the entropy function $s = s(e,n)$ is non analytic at the $\lambda$-line, and for our purposes this suffices for prescribing that the isothermal compressibility diverges there. Thus, this allows us to draw the isotherms as flat when crossing the $\lambda-$line. For the transition gas-superfluid we will also assume that such a line is one of critical points, and if so, in the same universality class as the $\lambda$-transition. This assumption is further validated by the behavior of the ideal Bose gas \cite{Buckingham} and by the recent experiments in ultracold gases, which show that the transition gas to superfluid is indeed second order \cite{Esslinger,Zwierlein}. We insist once more that it is not mandatory that the gas-superfluid be second order. If it were first order, the details of the phase diagrams would change, but not the main conclusions since these follow from the mere existence of the superfluid phase and not from the order of the phase transitions. We also point out that there are two triple points, a lower one connecting vapor, liquid and superfluid, and the upper one with solid, superfluid and liquid.\\

For the construction of the $p-n$ diagram, Fig. \ref{quantumall}(b), special care must be taken due to the existence of the superfluid phase. These states have the property of remaining fluid, with finite density, even at $T = 0$ ($s = 0$). From the thermodynamic point of view, the temperature $T = 0$ is unreachable because the zero isotherm and zero isentropic curves are degenerate and do not cross any other isotherm $T \ne 0$, neither any isentropic $s \ne 0$ curves. That is, it would take an infinite number of steps to reach the $T = 0$ state. But the state does ``exist": it is the ground state for the corresponding number of atoms. As a consequence of this, the isotherm $T = 0$ in the $p-n$ diagram, first of all, occurs for all (finite) values of $n$ and, second, it limits the existence of thermodynamic states to higher values of the corresponding pressure $p(n,T = 0)$. Thus, in Fig. \ref{quantumall}(b), the states in the shaded region below the $T = 0$ line are not ``unstable", it is simply meaningless to speak of them. As can also be seen from Fig. \ref{quantumall}(b), there are isotherms that can cross from the gas to the superfluid, then to the liquid and finally to the solid phase. These isotherms cross each other in the vicinity of the $\lambda$-line, namely, in the transition between liquid and superfluid. Such a crossing of different isotherms does not violate any law, it is simply a consequence that the slope of the $\lambda$-line is ``anomalous", namely, it is negative, just as in the solid-liquid transition line in water where the isotherms also cross. \\
 
 The crucial difference between a quantum and a normal substances appears in the construction of the $s-n$ diagram in Fig. \ref{quantumall}(c). In our search for the relationship $s = s(e,n)$, we need to draw curves at constant energy, $e =$ constant, in the $s-n$ diagram. Thus, the appearance of the superfluid, with states even as $s \to 0$ ($T \to 0$), makes it necessary that the curves $s$ vs $n$ at constant $e$, while always concave, necessarily are not monotonic. That is, the isoenergetic curves increase, initiating from $s = 0$ in the gas phase, then  reach a maximum, to finally decrease again to $s = 0$ in the superfluid or solid phases. This is best seen by considering a curve $s$ vs $n$ at very low energy in the vicinity of the transition gas to superfluid, see curve $e_i$ in Fig. \ref{quantumall}(c). As we have argued before, in the limit $n \to 0$ and $s \to 0$, the gas must approach an ideal gas. In such a case, the chemical potential is negative, $\mu <0$, namely, $s$ vs $n$ has a positive slope in the gas phase, see Eq. (\ref{Tmu}). On the other hand, at such a fixed energy, we know that there must be a number of particles for which the given energy equals their many-body ground state energy. Such a state is superfluid and has zero entropy. Since the curve must be concave, as the number of particles is slightly reduced from the $s = 0$ superfluid state, keeping the energy fixed, the entropy must increase and, therefore, the curve $s$ vs $n$ must have a negative slope in the vicinity of $s = 0$. That is, its chemical potential must be positive, $\mu >0$. This should not be surprising, Bogoliubov theory of atomic Bose superfluidity \cite{Bogoliubov} and BCS theory of atomic Fermi superfluidity \cite{BCS} both show positive chemical potentials near and at zero temperature. It is interesting to find that the ideal Fermi gas curves $s$ vs $n$ at $e = $ constant also all show a maximum and are all similar to the curve $e_i$ in Fig. \ref{quantumall}(c) \cite{VRR-arXiv}. The previous argument indicates that there exists a point where the curve $s$ vs $n$ at $e$ constant reaches a maximum, namely, when the chemical potential becomes zero, $\mu = 0$. \\
 
The final phase diagram $s-e$, shown in Fig. \ref{quantumall}(d), can be constructed from the $s-n$ diagram and taking into account that the isochores, $n = $ constant, $s$ vs $e$ must be monotonously concave since $T > 0$. The diagram $s-e$ appears extremely complex because the isoenergetic vertical lines fold over themselves. That is, there are no states above the line $\mu = 0$. This defines the locus of the local maxima of all isoenergetic curves. Nevertheless, with care and patience, one can find the main features shown in the diagram, in particular, one sees that the isochores curves cross each other. Again, this does not represent any problem since the single-valuedness of $s = s(e,n)$ is always ensured. That is, the multivaluedness of the function $n = n(s,e)$ does not violate any law since $e$ and $s$ are inverse functions of each other at $n =$ constant. In other words, the function $n = n(s,e)$ is not a fundamental relationship. \\

In Fig. \ref{s-e-n-cuantico} we show a condensed, yet simpler version of the surface $s = s(e,n)$. This figure show isentropic curves $s =$ constant in a $n-e$ diagram, similarly to  the diagram in Fig. \ref{DiagramaNVSENormal} of the normal substance. The main feature is that the isentropic curves are all acute, with their vertices or turning points at $\mu = 0$. This can be seen by analyzing the normal vectors $\hat n \sim (-1,\mu,T)$ to the surface $s = s(e,n)$, using Eq. (\ref{normal}) (with  $e$ along $x$, $n$ along $y$ and $s$ along $z$). One consequence is that adiabatic processes in these substances, being non-monotonic in density are also non monotonic in temperature. A second consequence is that the limiting curves $s = 0$ are both reached with infinite slope. Again, the appearance of a superfluid modifies profoundly the whole entropy surface with respect to a normal substance, as a direct comparison between Figs. \ref{DiagramaNVSENormal} and \ref{s-e-n-cuantico} show. \\

\begin{figure}[h]
\centering
\includegraphics[scale=0.5]{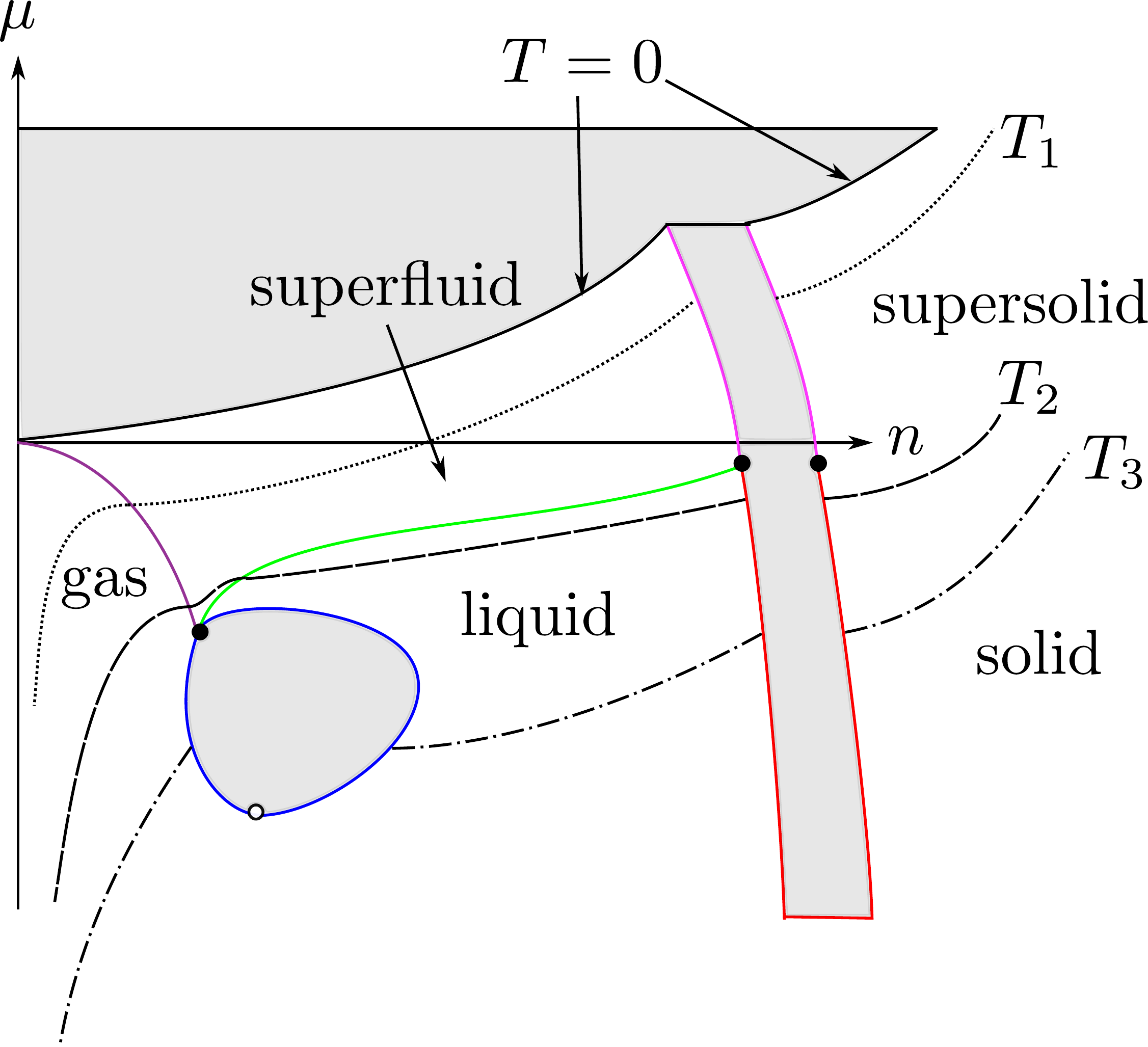}
\caption{(Color online) Isotherms in the $\mu-n$ projection for the quantum case. Here $T_1 < T_2 < T_3$. Note that we can distinguish a region in the solid phase with $\mu > 0$. This region is in coexistence with the superfluid and we could say that this is the supersolid phase of the quantum substance.} 
\label{mu-n-cuantico}
\end{figure}

Although there is no a priori reason to consider that the line of states with $\mu = 0$ represents a change in the thermodynamic properties, such as in a phase transition, it does appear to signal a kind of crossover between ``normal" and ``quantum" behaviors, as suggested in Section II. That is, if $\mu < 0$, the entropy increases with increasing density, at fixed temperature, a behavior that is in agreement with typical systems obeying classical mechanics: more particles, more entropy. However, in the extreme case in which systems can be arbitrarily close to $T = 0$, those systems are in quantum degenerate states, and it must be the opposite: increasing the number of particles reduces the number of states since the energy per particle is also reduced until the ground state is reached. Fig. \ref{mu-n-cuantico} shows a phase diagram $\mu-n$, where few isotherms are also plotted. The purpose of this diagram is to highlight the presence of states with positive chemical potential. The crossover line $\mu = 0$ appears to cross the superfluid region and, due to the observation of the previous paragraph, the actual existence of this line could be tested with adiabatic processes within the superfluid phase since it is only in this region where the non-monotonicity of the isentropic curves is present. That is, the liquid and gas region are all in the $\mu < 0$ region and, therefore, adiabatic processes are monotonic there. The other important aspect we want to highlight is that the solid region is divided into two regions, one with $\mu < 0$ and the other with $\mu > 0$. Because quantum properties are clearly more present in the $\mu > 0$, we want to hypothesize here that the solid region with $\mu > 0$ may be identified with the so-called ``supersolid" phase. As it has been experimentally shown, when superfluid $^4$He is solidified by increasing the pressure, at very low temperatures, the obtained solid has very peculiar mechanical properties, such as the mass transport through a cell filled with solid Helium \cite{Balibar,Hallock}. It has thus been speculated that this is a new phase, called supersolid. With the present study certainly we cannot predict all mechanical properties, however, we can tell that in adiabatic expansions  the energy is reduced instead of being increased, contrary to normal substances. Moreover, it must be clear that these solid states must be very different from the usual solid ones with $\mu < 0$, such as all the solids in a normal substance and the solid states at high temperature in quantum materials. That is, supersolids, just as superfluids, require a description using many-body quantum mechanics, that explicitly takes into account exchange effects due to the fact that the atoms are indistinguishable.  

\section{Final Remarks}

Thermodynamics methods and empirical evidence have made possible to obtain a qualitative geometric construction of the fundamental relationship $s = s(e,n)$ for both normal and quantum pure substances. We recall that, under the assumption that electromagnetic properties are not relevant for those systems, the fundamental relationship bears all their thermodynamic information and content. \\

We highlight that while the main property of the surface $s = s(e,n)$ is to be concave at each point, the existence of several phases indicates that the whole surface is in fact made of several ``patches" with each phase being covered by a smooth, continuous patch. First order phase transitions are actually the result of finite regions of unstable states that are actually not existent, with their borders defined by the continuity of the intensive quantities, temperature, pressure and chemical potential. Second order phase transitions occur either isolated, such as the liquid-vapor critical point, or are part of a finite set of states, such as the $\lambda-$line in the liquid-superfluid transition. In either case, those points are the ``suture" of two smooth surfaces and, therefore, this results on those states being non-analytic points of the surface $s = s(e,n)$.\\

A remarkable result, not easily foreseen, was the role of the sign of the chemical potential, which in turn just reflects the dependence of $s(e,n)$ on $n$ at $e =$ constant, assuming $T >0$ throughout. That is, negative chemical potential yields a concave {\it increasing} entropy $s$ as a function of $n$, while positive chemical potential a {\it decreasing} one. Our analysis indicate that normal substances, that do not show superfluid phases, have the as a signature $\mu < 0$ for all thermodynamic states. For quantum substances the conclusion is that there are regions of negative chemical potential, such as all gas and liquid phases, while the superfluid and solid ones can have both $\mu <0$ and $\mu > 0$. We have argued that as the quantum properties of the thermodynamic state become more important, the chemical potential becomes positive. The locus of $\mu = 0$ does not appear as a phase transition but rather as a crossover. The fact that the solid phase clearly gets divided by the $\mu = 0$ line (a local maximum of $s$ vs $n$ at $e = $constant) has prompted us to identify such a solid region as the recently discussed ``supersolid" phase. These regions certainly occur as the temperature is arbitrarily lowered. \\

We also remark again that the existence of superfluid phases, with its concomitant change of the sign of the chemical potential, globally affects the whole surface $s = s(e,n)$, making it obtuse for normal substances and acute for quantum ones. This has as a consequence that adiabatic processes in normal substances are always monotonic, while non-monotonic in the quantum ones. It is further interesting that the non-monotonicity cannot be experimentally tested in the gas and fluid phases but in the superfluid and solid phases only.\\

\end{document}